\documentclass[aps,prd,12pt,showpacs,notitlepage,nofootinbib,tightenlines]{revtex4-1}
\usepackage{amsmath}
\usepackage{bm}
\usepackage{times}
\usepackage{braket}
\usepackage{color}
\usepackage{epsfig}
\usepackage{slashed}
\usepackage{hyperref}
\usepackage{multirow}
\newcommand{\beq}{\begin{eqnarray}}
\newcommand{\eeq}{\end{eqnarray}}

\def \cpc{ {\bf Chin. Phys. C} }

\def \copc{ {\bf Comput. Phys. Commum. } }
\def \epjc{{\bf Eur. Phys. J. C} }

\def \jcap{ {\bf JCAP}  }
\def \npb{ {\bf Nucl. Phys. B} }
\def \plb{ {\bf Phys. Lett. B} }

\def \prt{  {\bf Phys. Rept.} }

\def \prd{ {\bf Phys. Rev. D} }
\def \prl{ {\bf Phys. Rev. Lett.}  }

\def \jhep{ {\bf JHEP}  }

\definecolor{Red}{rgb}{1.,0.,0.}

\definecolor{Blue}{rgb}{0.,0.,1.}

\definecolor{nicered}{rgb}{0.7,0.1,0.1}
\definecolor{nicegreen}{rgb}{0.1,0.5,0.1}
\def\lsim{ {\ \lower-1.2pt\vbox{\hbox{\rlap{$<$}\lower6pt\vbox{\hbox{$\sim$}}}}\ } }
\def\gsim{ {\ \lower-1.2pt\vbox{\hbox{\rlap{$>$}\lower6pt\vbox{\hbox{$\sim$}}}}\ } }

\hypersetup{colorlinks,citecolor=nicegreen,linkcolor=nicered}
\begin{document}
\title{Search for single production of the vector-like top partner at the 14 TeV LHC}
\author{Yao-Bei Liu\footnote{E-mail: liuyaobei@hist.edu.cn}, Yu-Qi Li}
\affiliation{Henan Institute of Science and Technology, Xinxiang 453003, P.R.China }
\begin{abstract}
The new heavy vector-like top partner~($T$) is one of typical
 features of many new physics models beyond the standard model.
 In this paper we study the discovery potential of the LHC for the vector-like $T$-quark
 both in the leptonic $T\to bW$ and $T\to t_{\rm lep}Z_{\rm lep}$ (trilepton) channels at $\sqrt{s}= 14$ TeV in the single production mode.
 Our analysis is based on a simplified model including a $SU(2)_L$ singlet with charge $2/3$ with
only two free parameters, namely the $TWb$ coupling parameter $g^{\ast}$ and the top partner mass
$m_T$. The $2\sigma$ exclusion limits, $3\sigma$ evidence and the $5\sigma$ discovery reach in the parameter
plane of $g^{\ast}-m_T$, are, respectively, obtained for some typical integrated luminosity at the 14 TeV LHC. Finally we analyze the projected sensitivity in terms of the production cross section times branching fraction for two decay channel.
\end{abstract}


\maketitle

\section{Introduction}
Various extensions of the standard model (SM) predict new heavy particles that address the
hierarchy problem caused by the quadratic divergences in the quantum-loop corrections
to the Higgs boson mass; for a review see~\cite{hmass}. The largest corrections, owing to the top-quark loop,
 are canceled by the existence of heavy partners of the top quark in many
of these models, such as little Higgs~\cite{littlehiggs}, extra dimensions~\cite{ED}
and composite Higgs~\cite{Agashe:2004rs} models.
The discovery of the 125 GeV Higgs boson~\cite{lhc-higgs}
at the Large Hadron Collider (LHC) has excluded heavy SM-like chiral fermions~\cite{prl2012:109-241802}.
 Therefore, we focus on the case of a $SU(2)_L$ singlet vector-like top partner ($T$).
  The effects on Higgs production and decay rates of loop diagrams including $T$ quarks are
  well below the precision of the current measurements~\cite{higgs-decay}.
  In many cases, the vector-like top partners mix with the SM quarks predominantly of
   the third generation and can also stabilize the electroweak vacuum~\cite{prd90-014007}.
The phenomenology of new heavy quarks has been widely studied in literature; see for example~\cite{p1,p2,p3,p4,p5,p6,p7,jhep-1304-004,NP,NP1,NP2} and the forthcoming direct searches at
the LHC will therefore play a important role in testing many models
predicting the existence of these states.

  The current combined results of ATLAS and CMS searches have established lower limits on mass of the vector-like top partners in the range of 550-900 GeV at center-of-mass energies
of 8 TeV ~\cite{atlas-cms-8} and 13 TeV~\cite{atlas-13}, depending on the assumed branching ratios. Most of the experimental searches assume
the top partners to be pair produced via the strong interaction, and these bounds strongly
depend on the assumptions on the decay
branching ratios and the properties of the top partner. Because the vector-like quarks can induce corrections to precisely measured observables of
the SM, the relevant model parameters
can also be constrained by the indirect searches of the electroweak precision
 observables~\cite{ew1,ew2,rb,prd-88-094010}. On the other hand, it is possible that the new
   vector-like top partners can
   significantly mix with the SM light quarks~\cite{bound1,bound2,Buchkremer:2013bha}. However, such indirect constraints on the mixing parameters may be relaxed if several multiplets
are present in the low-energy spectrum~\cite{CHM}.
For high $m_T$ (about $m_T \gsim 1$ TeV), previous study showed that single production of top partners starts to dominate over pair production
  due to larger phase space~\cite{s1,s2,prd-92-011701,prd-90-075009,prd-90-115008,jhep-1311-047,jhep-2015-01-088,jhep-1604-014}. Especially for the case that the top partners can mix in a sizable
way with lighter quarks, their production cross section will be very
large due to the mixing with valence quarks~\cite{jhep-1108-080,jhep-1502-032,lyb-2017}.
We do not consider this case because the masses of top partners are not connected to electroweak symmetry breaking.

Very recently, Both the ATLAS~\cite{a-13} and the CMS~\cite{c-13}
 Collaborations presented a search optimized for a single produced
 vector-like $T$ quark at $\sqrt{s}=13$ TeV, subsequently decaying
  as $T\to Wb$ with leptonic decays of the $W$ boson. The smallest coupling limits on the $TWb$ coupling strength has been set as $|c_{L}^{Wb}|=0.45$ for a vector-like top partner with a mass of 1 TeV~\cite{a-13}. This encouraged us to further analyze this process in order to provide an effective search strategy for the future 14 TeV LHC. In particular, we also studied the observational potential of single vector-like top partner in the $T\to t_{\rm lep}Z_{\rm lep}$ decay channel at 14 TeV high-luminosity (HL)-LHC with an
integrated luminosity of 3 ab$^{-1}$. Although the branching ratio of $T\to t_{\rm lep}Z_{\rm lep}$~(about $0.37\%$) decay channel is small and results in a suppressed production rate for the final states, it has the
great advantage of small QCD backgrounds~\cite{jhep-1411-104,14051617}.
Therefore, we here mainly study the observability of a single $T$-quark
 production at the 14 TeV LHC both for the leptonic $T\to bW$ and for the $T\to t_{\rm lep}Z_{\rm lep}$ (trilepton) channels, and we discuss the event selection and cuts on kinematic variables in detail. Finally, the exclusion limits and discovery potential of the production cross section times branching fraction for two decay channels are, respectively, examined as a function of top partner mass for several typical luminosity at the LHC. In order to keep
  the model as independent as possible, we here perform the study in the
   framework of a simplified model, which only comprises two independent parameters.

The rest of the paper is organized as follows. In Sec.~II we briefly describe
the main features of the simplified model. In Sec.~III we turn to study the
 prospects of observing the single $T$ production by performing a detailed
 analysis of the signal and backgrounds in both the leptonic $Wb$ and the $T\to t_{\rm lep}Z_{\rm lep}$ decay channels at 14 TeV LHC.
 Finally, we conclude in Sec.~IV.

\section{Top partner in the simplified model}
As mentioned above, the benefit of using of simplified models is that the results of the studies could be used to make predictions for more complex
models including the top-partners. A generic parametrization of an effective Lagrangian for top partners has been proposed in
Ref.~\cite{Buchkremer:2013bha}, where the vector-like quarks are embedded in different representations
 of the weak $SU(2)$ group. We here consider a simplified model where the vector-like $T$-quark is
 an $SU(2)$ singlet with charge $2/3$, with couplings only to the third generation of SM quarks.
The benefit of using the simplified effective theory  is that the results of the studies could be
used to make predictions for more complex models including various types of top partners.

 The top-partner sector of the model is described by the general effective Lagrangian
  (showing only the couplings relevant for our analysis)~\cite{Buchkremer:2013bha}
\begin{eqnarray}
{\cal L}_{T} =&& \frac{g^{\ast}}{\sqrt{2}}[ \frac{g}{\sqrt{2}} \bar{T}_{L}W_{\mu}^{+}
    \gamma^{\mu} b_{L}+
    \frac{g}{2\cos \theta_W}\bar{T}_{L} Z_{\mu} \gamma^{\mu} t_{L}
    - \frac{m_{T}}{v}\bar{T}_{R}ht_{L} -\frac{m_{t}}{v} \bar{T}_{L}ht_{R} ]+ h.c. ,
  \label{TsingletVL}
\end{eqnarray}
where $m_T$ is the top partner mass, and $g^{\ast}$ parametrizes the single production
 coupling in association with a $b$- or a top-quark. $g$ is the $SU(2)_L$ gauge coupling
  constant, $v\simeq 246$ GeV and $\theta_W$ is the Weinberg angle. Thus there are only two model parameters: the top partner mass $m_T$ and the coupling strength to SM quarks in units of standard couplings, $g^{\ast}$. Here we take a
   conservative range for the coupling parameter~\cite{a-13,vtb}:
$g^{\ast}\leq 0.5$, which is consistent with the current experiment bounds.

\begin{figure}[htb]
\begin{center}
\vspace{0.5cm}
\centerline{\epsfxsize=12cm \epsffile{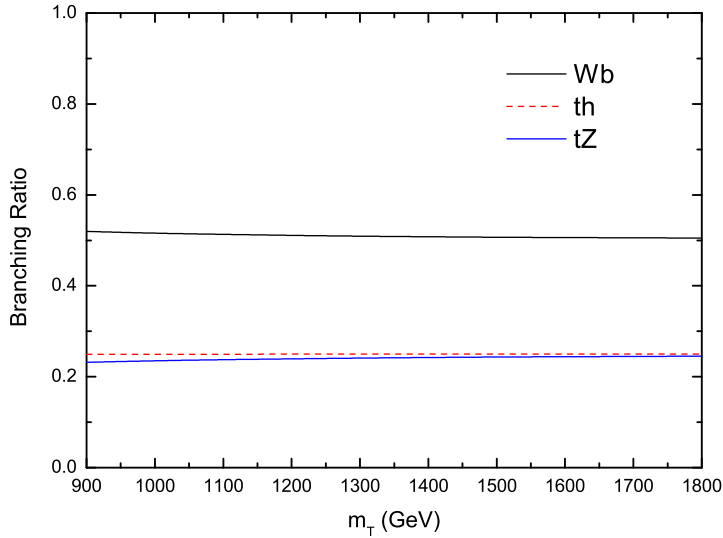}}
\caption{Branching ratios for the three decay modes for various of top partner masses
with $g^{\ast}=0.2$}
\label{br}
\end{center}
\end{figure}

In general, the vector-like $T$-quark has three different decay
 channels into SM particles: $bW$, $tZ$, and $th$. In Fig.~\ref{br}, we show the branching ratios
  of three decay channels by varying top partner
 masses with $g^{\ast}=0.2$.
 We can see that $Br(T\to th)\approx Br(T\to tZ)\approx \frac{1}{2}Br(T\to Wb)$
 is a good approximation as expected by the Goldstone
boson equivalence theorem~\cite{ET-prd}.

\section{Event generation and analysis}
\begin{figure}[htb]\vspace{-1.5cm}
\begin{center}
\centerline{\epsfxsize=20cm \epsffile{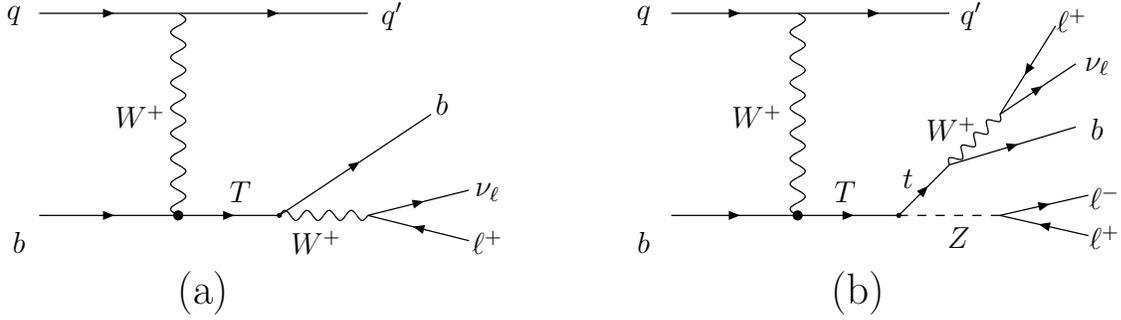}}
\vspace{-21cm}
\caption{The Feynman diagram for production of single $T$ quark including the decay chains $T\to bW(\to \ell^{+} \nu)$ and $T\to t(\to b\ell^{+} \nu)Z(\to \ell^{+}\ell^{-})$ }
\label{ppT}
\end{center}
\end{figure}

In this section, we analyze the observation potential by performing a
Monte Carlo simulation of the signal and background events and explore
 the sensitivity of single top partner at the LHC through $T\to bW$ and $T\to tZ$ channels.
The Feynman diagram of the
production and decay chain is presented in Fig.~\ref{ppT}.

The model file generating signal events according to the Lagrangian
of Eq.(\ref{TsingletVL}) can be found in the dedicated FeynRules~\cite{feynrules} model
 database webpage~\cite{web}. The SM input parameters relevant in our study are
 taken from  \cite{pdg}. The corresponding free parameters are the top-partner mass $m_T$
 and the coupling parameter $g^{\ast}$ which governs the top-partner single production
 involving a $t$-channel $W$ boson. Considering the current constraints from the top-partners pair
 production processes at 13 TeV ATLAS detector~\cite{atlas-13}, we generate eight benchmark points
 varying the $T$ mass
 in steps of 100 GeV in the range
$m_T\in [900; 1800]$ GeV with $g^{\ast}=0.2$. The QCD next-to-leading order~(NLO)
prediction for the single top partners productions are calculated in
Refs.~\cite{jhep-2015-01-088,1610.04622}.
 Following Ref.~\cite{1610.04622}, we here take the conservative value of the $K$-factor
  as 1.2 for the signal before the event generation.

Signal and background events are simulated at the leading order using
MadGraph5-aMC$@$NLO~\cite{mg5} with the CTEQ6L parton distribution function (PDF)~\cite{cteq}
 and the renormalization and factorization scales are set dynamically by default.
 Pythia6~\cite{pythia} and Delphes~\cite{delphes} are used to perform the parton
shower and the fast detector
simulations, respectively. The anti-$k_{t}$ algorithm
\cite{antikt} with parameter $\Delta R=0.4$ is used to reconstruct the jets.
Event analysis is performed by using the program of MadAnalysis5 \cite{ma5}.

\subsection{The $T\to Wb$ channel}
In this section, we analyze the observation potential by performing a
Monte Carlo simulation of the signal and background events and explore the sensitivity of single top partner at the LHC through the channel
\beq\label{signal}
pp \to T(\to bW^{+})j\to bW^{+}( \to \ell^{+}\bar{\nu}_{\ell}) j.
\eeq

For this channel, the typical signal is exactly one charged lepton, one $b$ jet, one forward jet and missing energy.
The dominant background turns out to be the $W+$ light jets with one of the jets
 misidentified as $b$-quark jet and $t\bar{t}$~(semi-leptonic) + jets. $W+b+$
  light jets and $W+b\bar{b}$ can also make contribution the backgrounds.
Meanwhile, the $t\bar{t}$ samples are normalised to the theoretical cross-section
    value for the inclusive $t\bar{t}$ process of 953.6 pb performed at
    next-to-next-to-leading order (NNLO) in QCD and including resummation
    of next-to-next-to-leading logarithmic (NNLL) soft gluon terms~\cite{1303.6254}.
    The QCD corrections for the
dominant backgrounds are considered by including a $k$ factor, which is 1.12 for $W^{+}bj$
\cite{nlo-wbj}, 1.5 for $W^{+}b\bar{b}$ \cite{nlo-wbj},
and about 1.2 for $W^{+}jj$ \cite{nlo-wjj,14050301}.
On the other hand, the MLM matching scheme is used, where we included up to three extra jets
    for $W_{\rm lep}$ + jets and
up to one additional jet to $t\bar{t}$ in the simulations~\cite{MLM}.
Other smaller backgrounds come from single top ($tW$, $t$-channel and $s$-channel
 with up to one additional jet) and diboson~($WW$, $ZZ$, $WZ$) production.
 The cross sections are scaled according to the approximate NNLO theoretical
  predictions~\cite{nnlo-ww,nnlo-tw,nnlo-tj}.

In our simulation, all signal and background events are required to pass the following basic cuts:
\begin{itemize}
\item
There is exactly one isolated electron or muon ($N_{\ell}=1$) with $p_{T}^{\ell} > 25 \rm ~GeV$
 and $|\eta_{\ell}|<2.5$.
\item
Jets are required to satisfy $p_{T}^{b} > 25 \rm ~GeV$ and $|\eta_{b}|<5.0$.
 There are exactly one b-tagged jet ($N_{b}=1$) with $p_{T}^{b} > 25 \rm ~GeV$
 and $|\eta_{b}|<2.5$ and there are no more than three jets in total ($N_{j}<3$).
\item
The missing transverse momentum $\slashed E_T^{miss}$ is required to be larger than 20 GeV.
\end{itemize}

\begin{figure}[htb]
\begin{center}
\vspace{0.5cm}
\centerline{\epsfxsize=13cm \epsffile{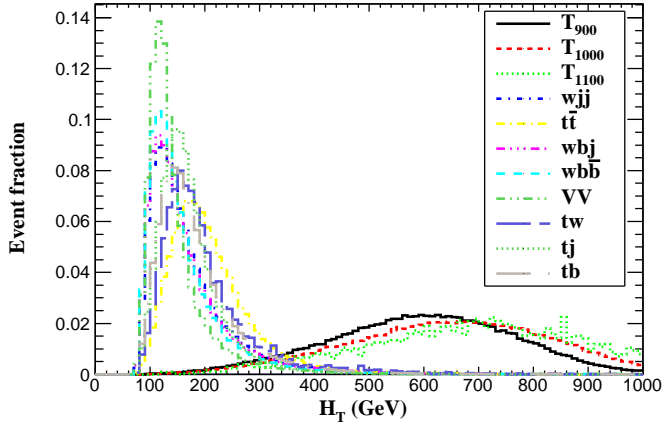}}
\caption{Normalized scalar sum of the transverse momenta $H_T$ for the signals and backgrounds}
\label{HT}
\end{center}
\end{figure}
\begin{figure}[htb]
\begin{center}
\centerline{\epsfxsize=9cm\epsffile{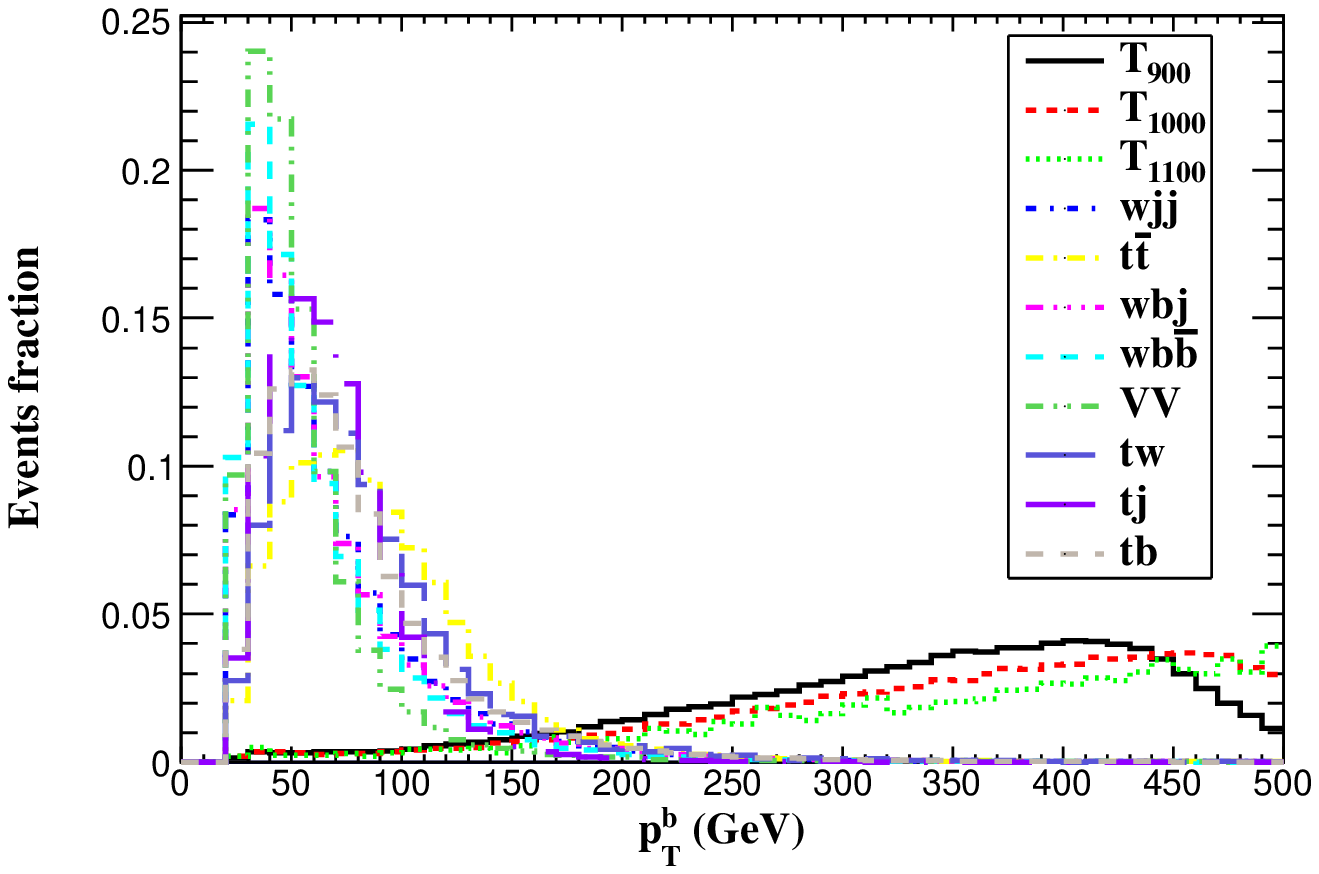}
\hspace{-1.0cm}\epsfxsize=9cm\epsffile{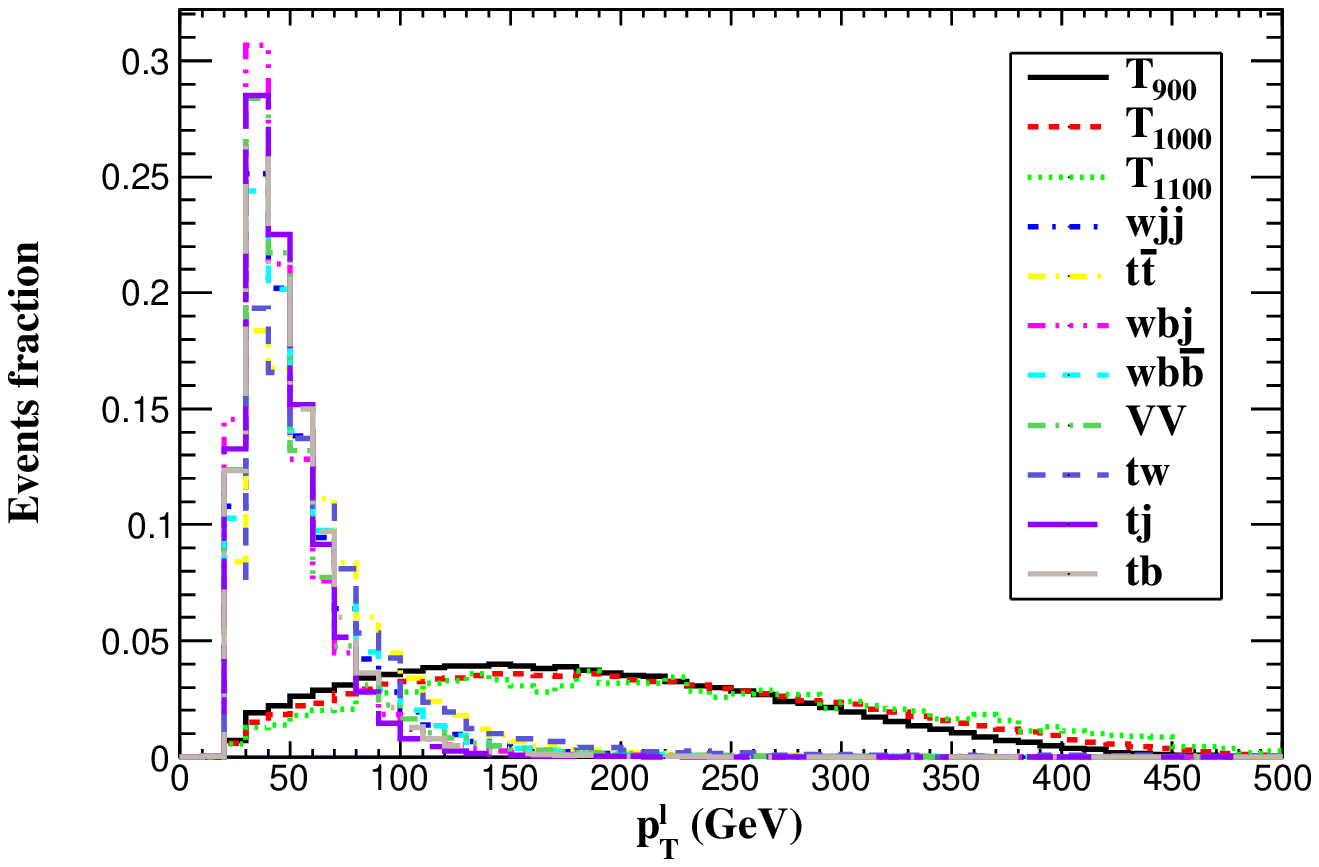}}
\centerline{\epsfxsize=9cm\epsffile{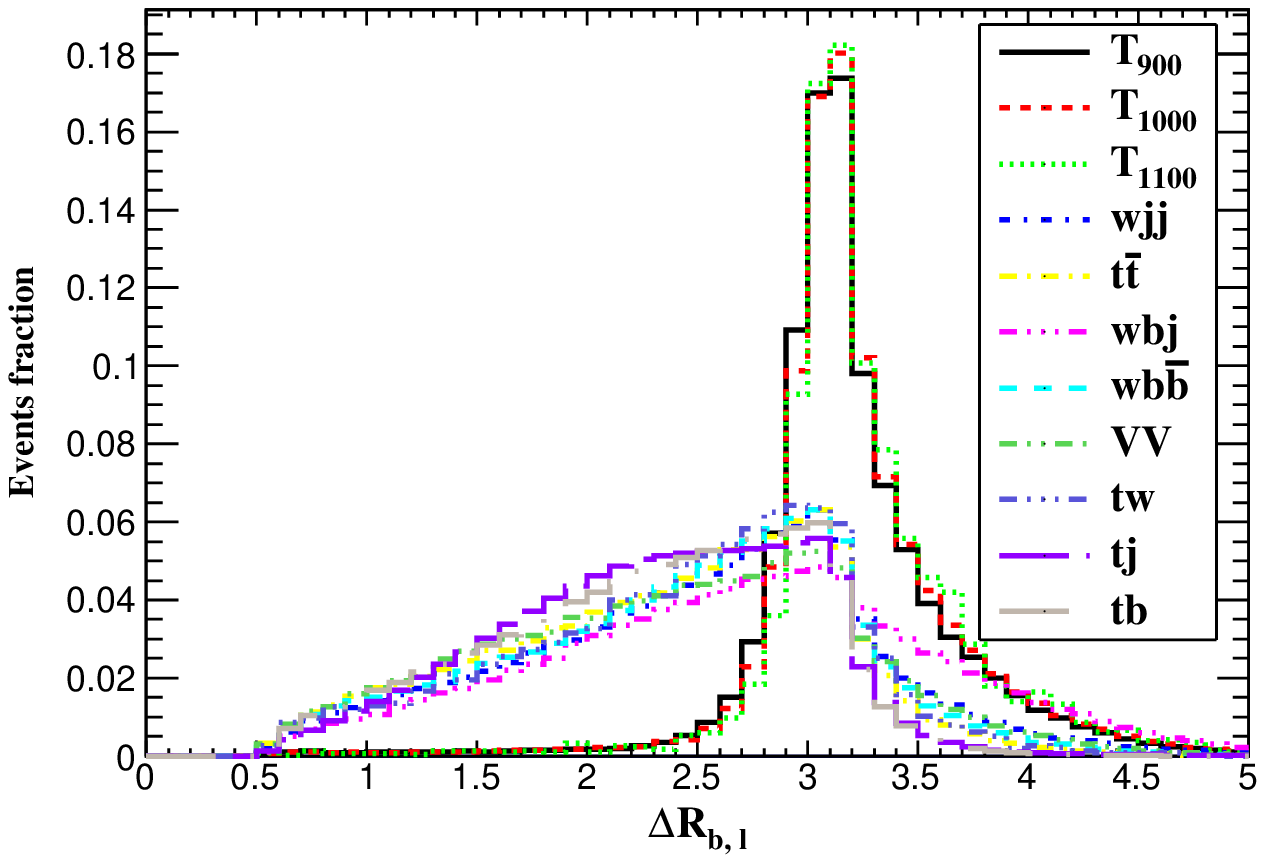}
\hspace{-1.0cm}\epsfxsize=9cm\epsffile{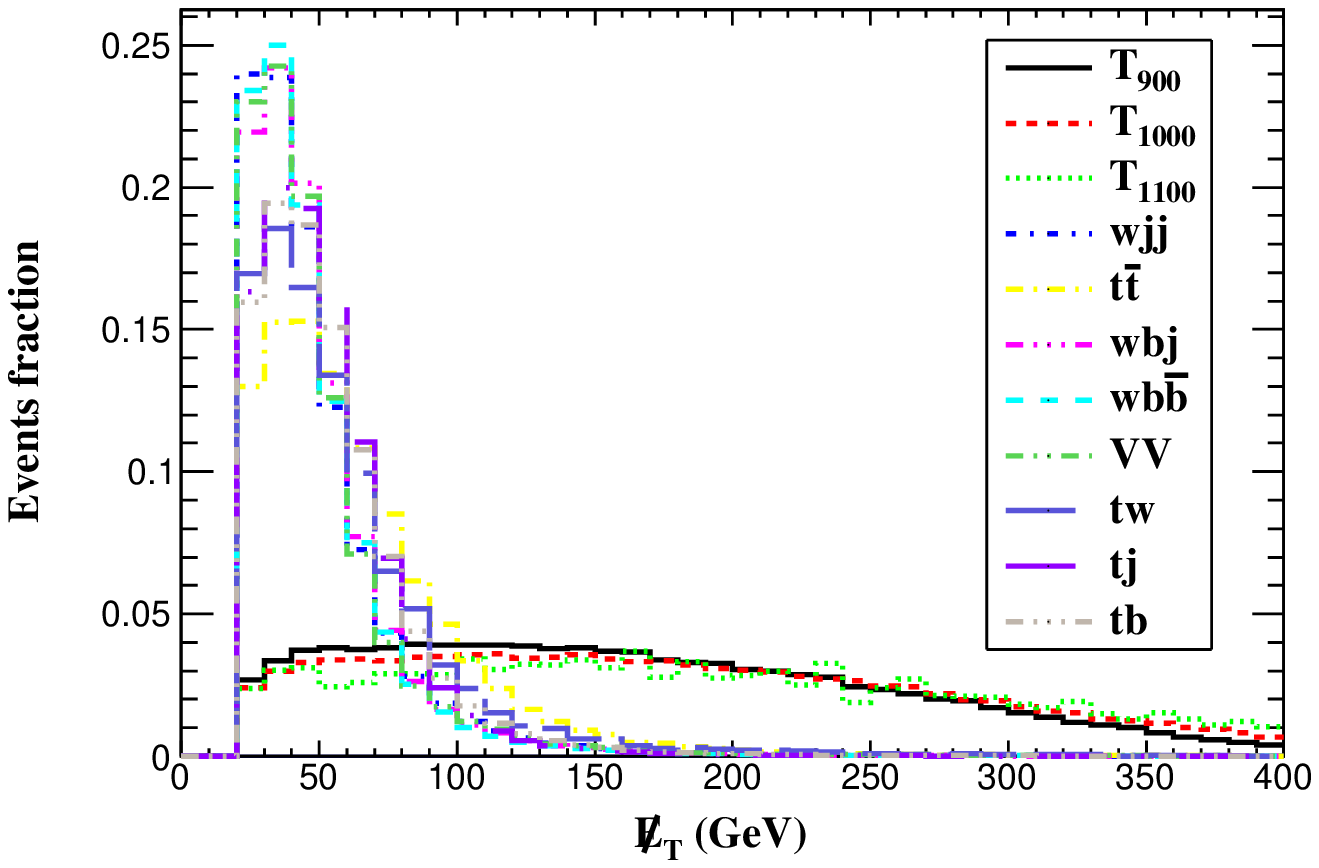}}
\caption{Normalized distributions of the transverse momenta ($p_{T}^{b}$ and $p_{T}^{\ell}$),
 $\Delta R_{b,\ell}$ and $\slashed E_T$ for the signals and backgrounds}
\label{wb}
\end{center}
\end{figure}

In order to choose appropriate kinematic cuts, we show some important kinematic distributions for
 the signal and the backgrounds. In Fig.~\ref{HT}, we show the normalized distribution of the
signals and backgrounds on $H_T$, defined as the scalar sum of the transverse momenta of
the $b$-tagged jet, the untagged jet and the lepton. From the figure, we can see that the
distributions of $Wjj$, $t\bar{t}$, single top and diboson backgrounds have peaks below 200 GeV,
while the peak position of the signals are larger than 500 GeV. Thus we choose
the $H_T$ cut as follows.
\begin{itemize}
\item Cut 1: $H_T>500 \rm GeV$.
\end{itemize}

In Fig.~\ref{wb}, we show the normalized distributions of the transverse momenta
$p_{T}^{\ell}$, $p_{T}^{b}$, the variable $\Delta R(b,\ell)$ and the missing transverse momentum
$\slashed E_T$ for the signals and backgrounds. Here $\Delta R=\sqrt{(\Delta\phi)^{2}+(\Delta\eta)^{2}}$
 is the particle separation among the objects (the tagged $b$-jet and
the lepton) in the final state with $\Delta\phi$ and $\Delta\eta$ being the separation in the
azimuth angle and rapidity, respectively. In the decay of
a singly produced top partner, the lepton from the leptonic $W$ boson decay and the $b$ quark tend to be produced with the transverse
momenta pointing in opposite directions. On the other hand, since the $W$ boson originating from heavy top partner decay has significant transverse momentum $p_T$, events are required to have large missing energy due to the undetected neutrino from the $W$ boson decay. Based on these kinematical distributions, we impose
 the following cuts to
get a high significance.
\begin{itemize}
\item Cut 2: $p_{T}^{\ell}> 100 \rm ~GeV$, $p_{T}^{b}> 250 \rm ~GeV$, $2.8 < \Delta R(b,\ell)< 3.5$
 and $\slashed E_T> 100 \rm ~GeV$.
\end{itemize}

\begin{figure}[htb]
\begin{center}
\vspace{0.5cm}
\centerline{\epsfxsize=13cm \epsffile{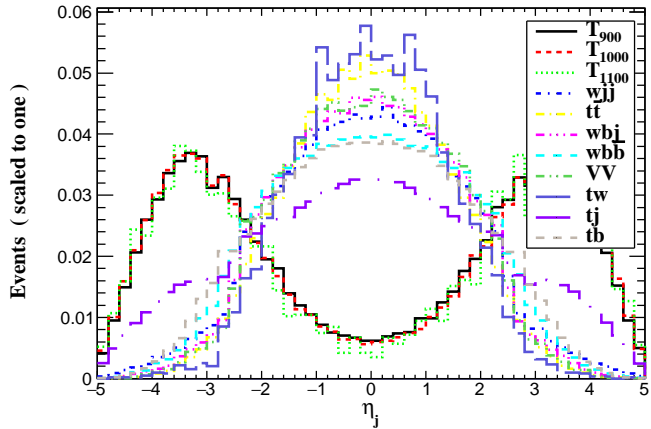}}
\caption{Normalized distribution of the rapidity of the forward jet for the signals and backgrounds}
\label{etaj}
\end{center}
\end{figure}

Since the jet from splitting of a valence quark with one $W$ emission always has a strong
forward nature, we plot the distribution of the rapidity of the forward jet in Fig.~\ref{etaj}
for the signals and backgrounds. From these distributions we can efficiently reduce the
 backgrounds by requiring the light untagged jet to have the following cut.
\begin{itemize}
\item Cut 3: $\mid\eta_{j}\mid > 2.4$.
\end{itemize}

\begin{figure}[htb]
\begin{center}
\vspace{0.5cm}
\centerline{\epsfxsize=13cm \epsffile{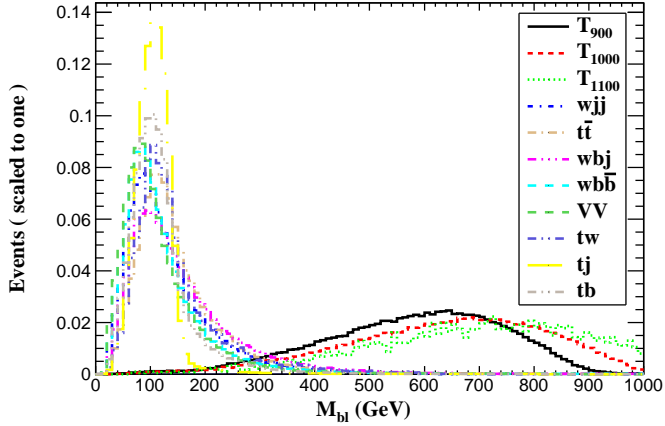}}
\caption{Normalized invariant mass distribution of $b\ell$ system for the signals and backgrounds}
\label{mbl}
\end{center}
\end{figure}

The invariant mass of the $b$-tagged jet and the lepton is plotted in Fig.~\ref{mbl} for the signals
 and the backgrounds.  One can see that the signal distributions
peak close to the $T$-quark mass, while background distributions
turn over at lower masses. Thus we can further reduce the backgrounds by the following cut.
\begin{itemize}
\item Cut 4: $M_{b\ell}>500 \rm~GeV$.
\end{itemize}

\begin{table}[htb]
\begin{center}
\caption{The cut flow of the cross sections (in fb) for three typical signals and the relevant
backgrounds at the 14 TeV LHC. Here we take the parameter as $g^{\ast}=0.2$ \label{cutflow1}}
\vspace{0.2cm}
\begin{tabular}{|c|c|c|c|c|c|c|c|c|c|c|c|}
\hline
\multirow{2}{*}{Cuts}& \multicolumn{3}{c|}{Signal}&\multirow{2}{*}{$W^{+}jj$}
 &\multirow{2}{*}{$t\bar{t}$} &\multirow{2}{*}{$W^{+}bj$} &\multirow{2}{*}{$W^{+}b\bar{b}$}
  &\multirow{2}{*}{$VV$} &\multirow{2}{*}{$tW$} &\multirow{2}{*}{$tj$} &\multirow{2}{*}{$t\bar{b}$}\\ \cline{2-4}
 &{900~GeV} &{1000~GeV}&{1100~GeV}&&&&&&&& \\  \cline{1-12}
\hline
Basic cuts& 2.84&1.96&1.58&2012&106&238&361&27&97&3066&166 \\ \hline
Cut 1&1.96&1.48&1.16&17.1&1.7&2.2&2.1&0.1&1.84&7.7&1.3\\ \hline
Cut 2&1.24&0.92&0.72&2.19&0.37&0.12&0.23&0.02&0.27&0.09&0.14\\ \hline
Cut 3&0.88&0.68&0.52&0.4&0.015&0.015&0.04&0.004&0.027&0.012&0.023\\ \hline
Cut 4&0.76&0.64&0.48&0.31&0.011&0.012&0.036&0.002&0.024&0.006&0.013\\ \hline
\end{tabular} \end{center}\end{table}

We present the cross sections of three typical signal ($m_T=900, 1000, 1100$ GeV) and the relevant
backgrounds after imposing
the cuts in Table~\ref{cutflow1}.
From Table~\ref{cutflow1},
one can see that all the backgrounds are suppressed very efficiently after imposing the selections.
From the numerical results as listed in the last line of Table~\ref{cutflow1},
one can see that $W^{+}jj$ is the most dominant background after applying all those
mentioned cuts.
To estimate the observability quantitatively, we adopt the significance measurement~\cite{ss}:
\beq
SS=\sqrt{2\pounds_{int}[(\sigma_S+\sigma_B)\ln(1+\sigma_S/\sigma_B)-\sigma_S]},
\eeq
where $\sigma_S$ and $\sigma_B$ are the signal and background cross sections
 and $\pounds_{int}$ is the integrated luminosity.
Here we define the discovery significance as $SS=5$, the possible evidence as $SS=3$ and the exclusion limits as $SS=2$. We do not consider the theoretical and systematic uncertainties, such as the choice of PDF set, the renormalization and factorization scales and the respective normalisation to the theoretical NNLO cross-sections, but we expect this will not change our results significantly.

\begin{figure}[htb]
\begin{center}
\centerline{\epsfxsize=9cm \epsffile{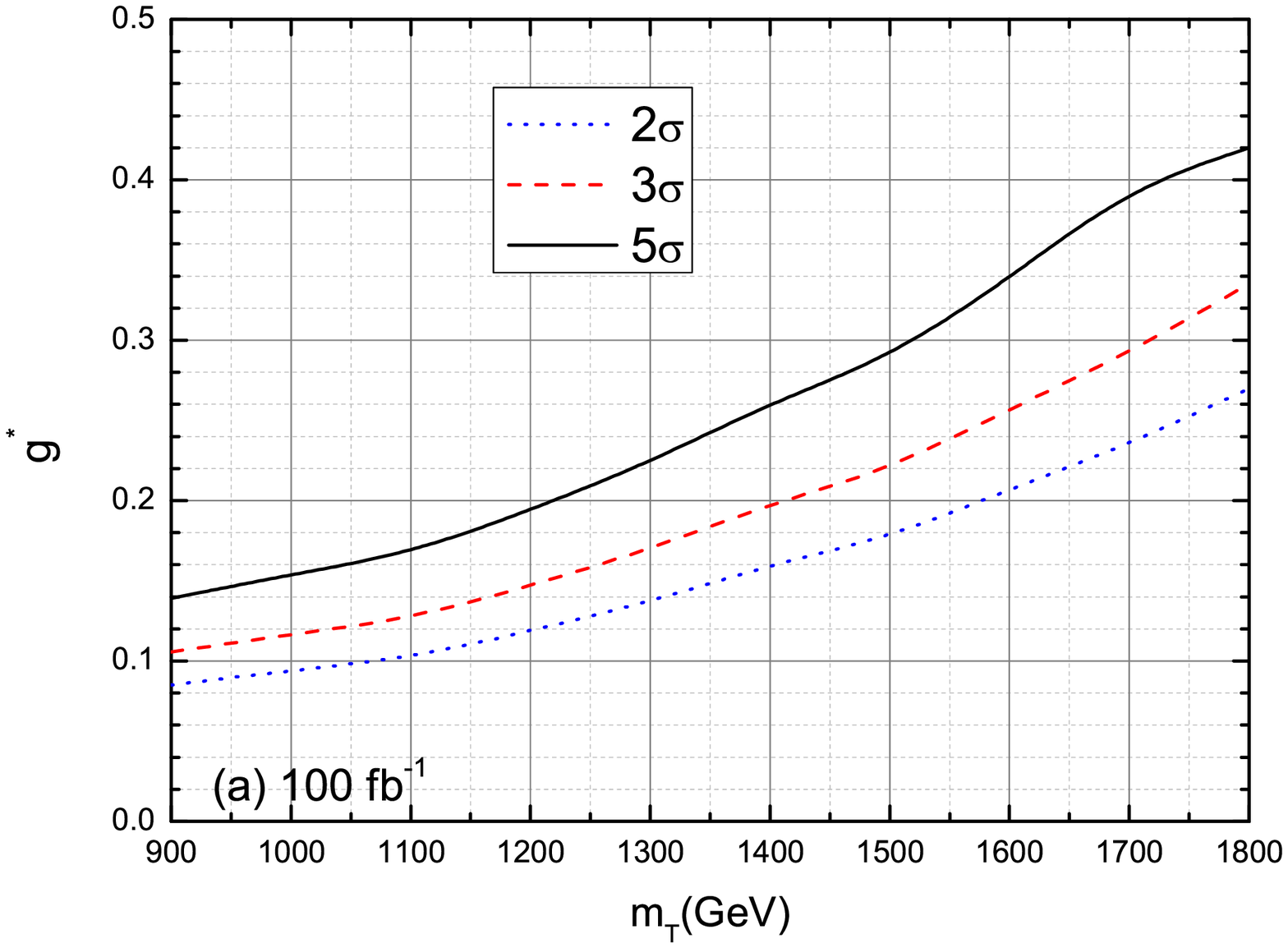}\epsfxsize=9cm \epsffile{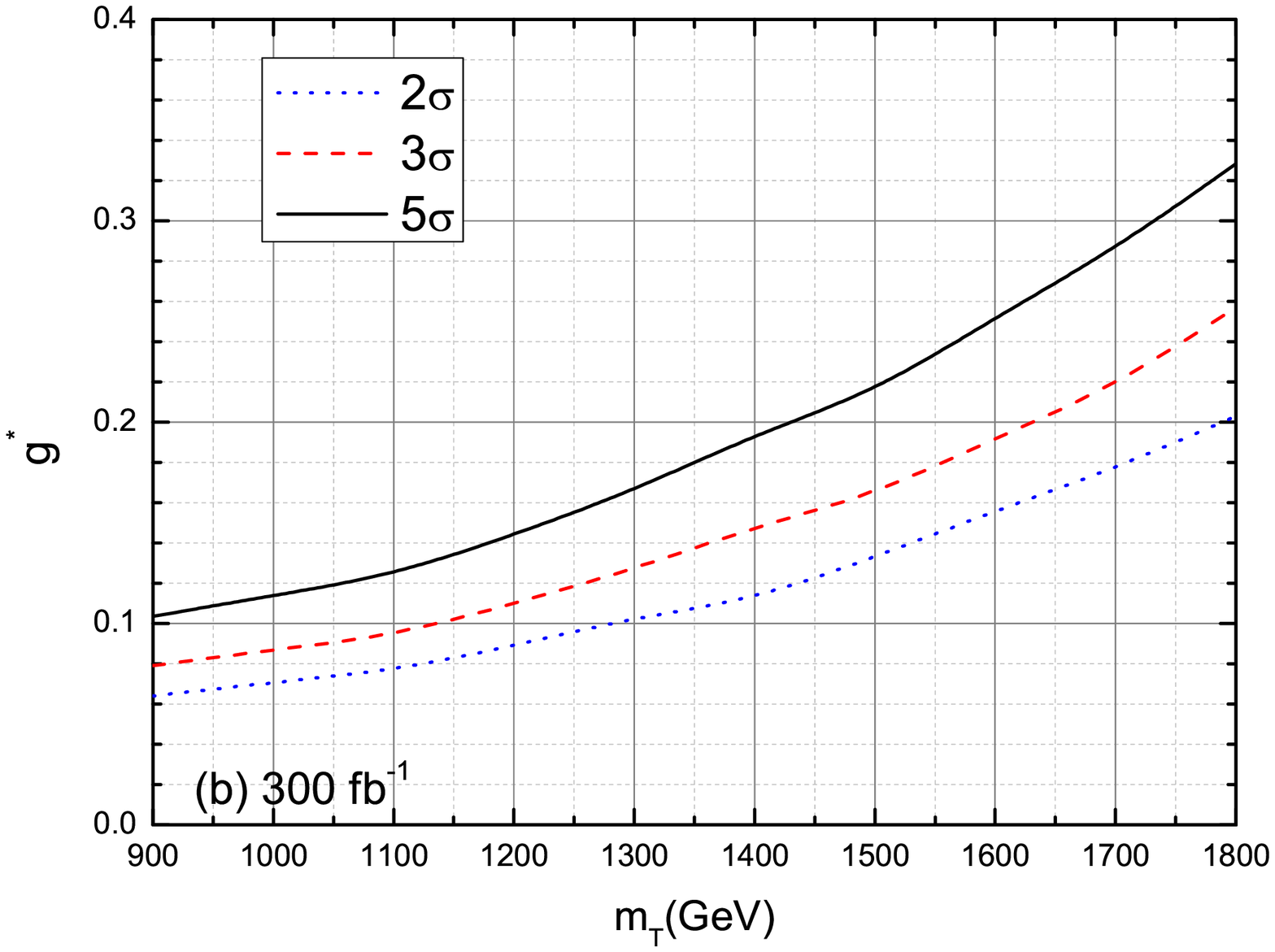}}
\caption{The $2\sigma$, $3\sigma$ and $5\sigma$ contour plots for the signal in $g^{\ast}-m_T$ at 14 TeV
LHC with two typical values of integrated luminosity: (a) 100 fb$^{-1}$, (b) 300 fb$^{-1}$.. }
\label{ss-wb}
\end{center}
\end{figure}

In Fig.~\ref{ss-wb}, the $2\sigma$, $3\sigma$ and $5\sigma$ lines are drawn as a function of
$g^{\ast}$ and the top partner mass $m_T$ for two fixed values of integrated
luminosity: 100 and 300 fb$^{-1}$. We can see that, for $m_T=0.9~(1.6)$ TeV,
the $5\sigma$ level discovery sensitivities of $g^{\ast}$ are, respectively, about
 0.14~(0.34) with $\pounds_{int}=100$ fb$^{-1}$
  and 0.10~(0.22) with $\pounds_{int}=300$ fb$^{-1}$. On the other hand, from the $2\sigma$
   exclusion limits one can see that for, $m_T=1.0~(1.5)$ TeV, the upper limits on the size
    of $g^{\ast}$ are, respectively, given as $g^{\ast}\leq 0.12~(0.22)$  with $\pounds_{int}=100$ fb$^{-1}$
     and $g^{\ast}\leq 0.09~(0.16)$ with $\pounds_{int}=300$ fb$^{-1}$.

\subsection{The $T\to tZ$ channel}
Next, we analyze the observation potential and explore the sensitivity of single $T$-quark at the 14 TeV LHC through the channel
\beq\label{signal}
pp \to T(\to tZ)j\to t( \to b\ell^{+}\bar{\nu}_{\ell})Z(\to \ell^{+}\ell^{-}) j.
\eeq

The main SM backgrounds that can give three leptons in the final state are $t\bar{t}$, $WZjj$, $t\bar{t}V(V=Z/W)$ and the irreducible $tZj$. In the $t\bar{t}$ case (both top quarks decay semi-leptonically), a third lepton comes from a semi-leptonic B-hadron decay in the b-jet. Here we do not consider multijet backgrounds where jets can be faked as electrons since they are very negligible in multileptons analyses~\cite{14067830}.

First of all, we apply the following cuts on the signal and background events.
\begin{itemize}
\item Basic cuts: $p_{T}^{\ell, b} > 25 \rm ~GeV$, $p_{T}^{j} > 40 \rm ~GeV$, $|\eta_{\ell, b}|<2.5$, $|\eta_{j}|<5$,  where $\ell=e, \mu$.
\end{itemize}
 Further, we apply some general preselections as follows.
 \begin{itemize}
\item Cut-1: There are exactly three isolated leptons ($N(\ell)\equiv3$), at least two jets and no more than three ($2\leq N(j) \leq 3$), of which exactly one is b-tagged ($N(b)\equiv 1$).
\end{itemize}
 The requirement of three leptons can strongly reduce the $t\bar{t}$ backgrounds, and the b-tagging can efficiently suppress the diboson components.

\begin{figure}[htb]
\begin{center}
\centerline{\epsfxsize=9cm\epsffile{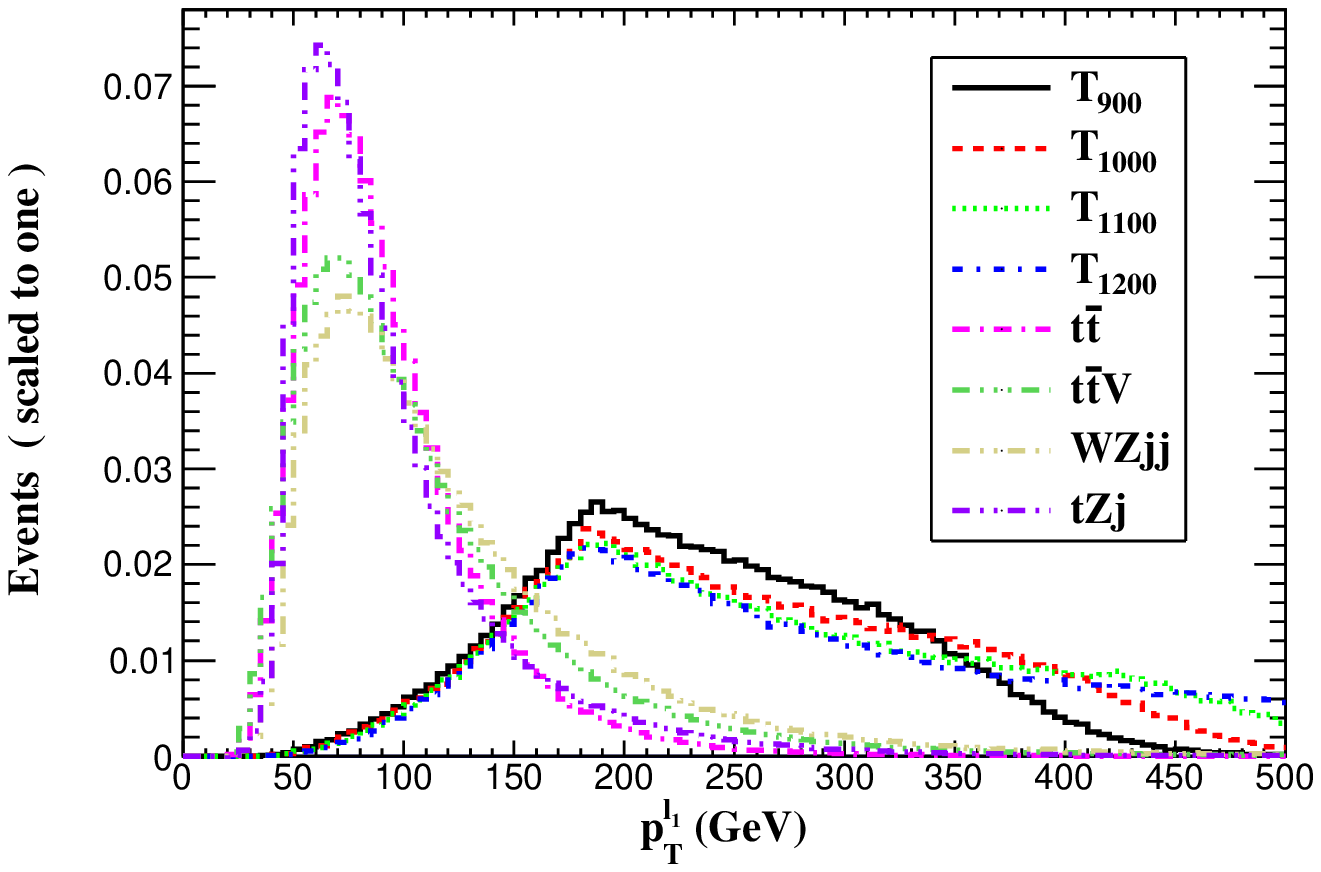}
\hspace{-1.0cm}\epsfxsize=9cm\epsffile{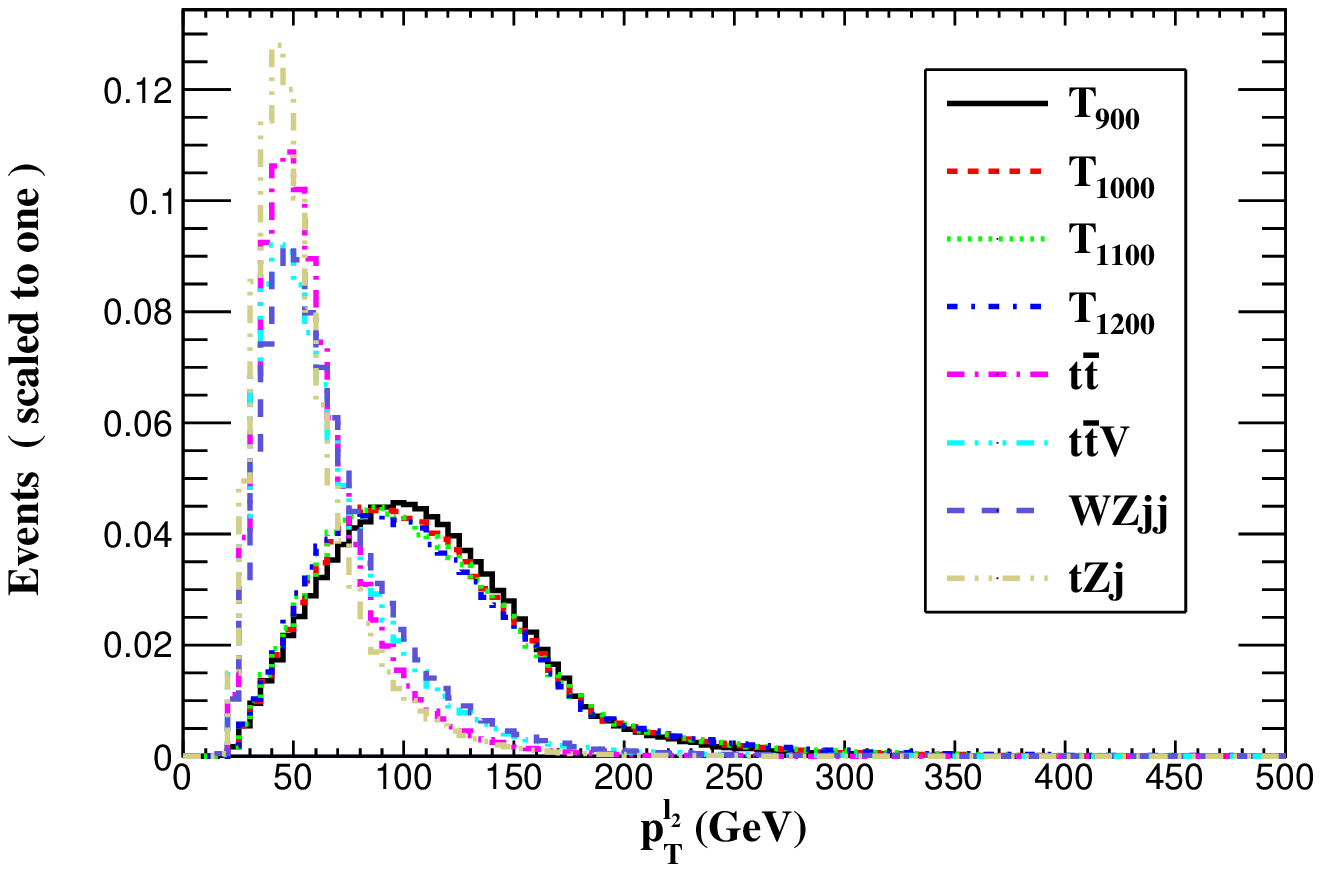}}
\centerline{\epsfxsize=9cm\epsffile{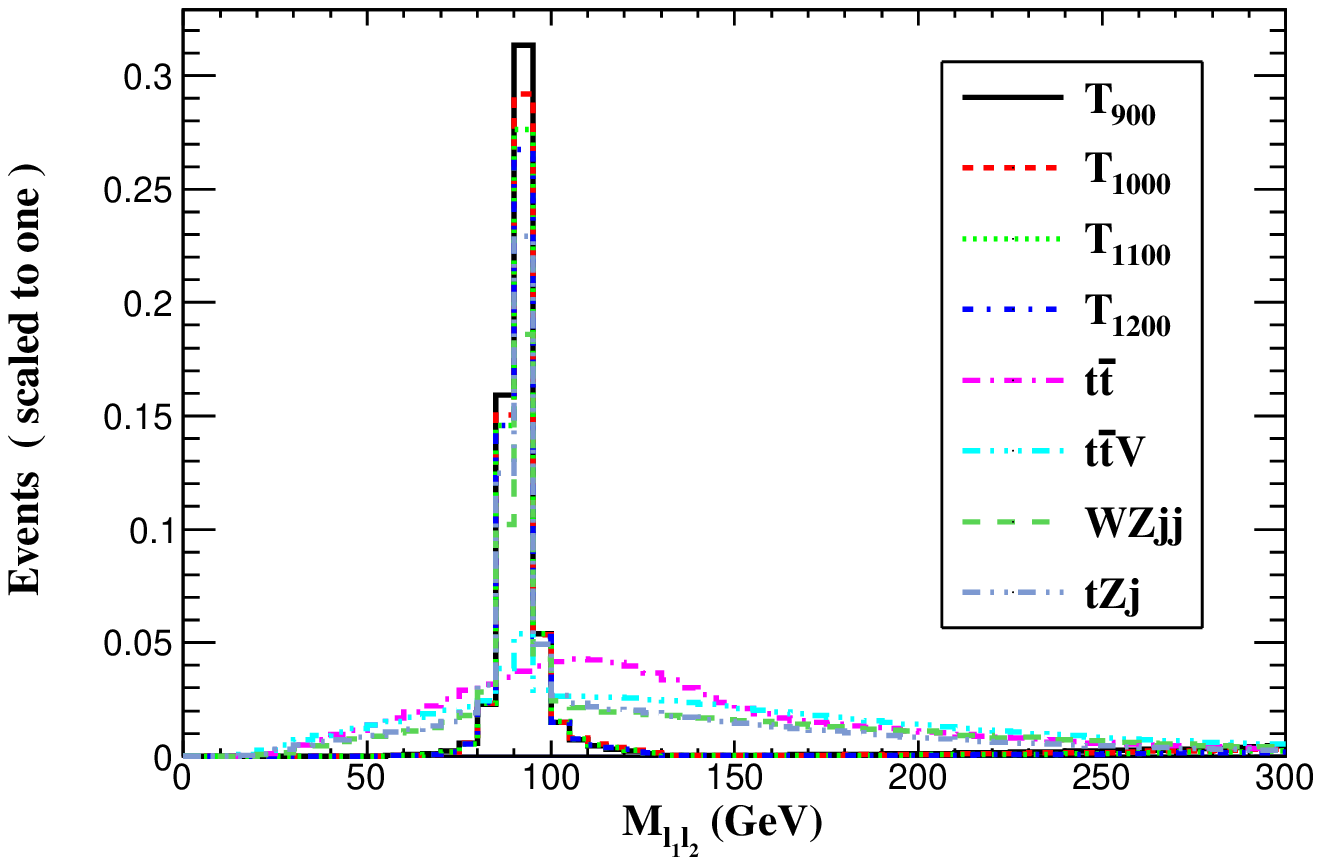}
\hspace{-1.0cm}\epsfxsize=9cm\epsffile{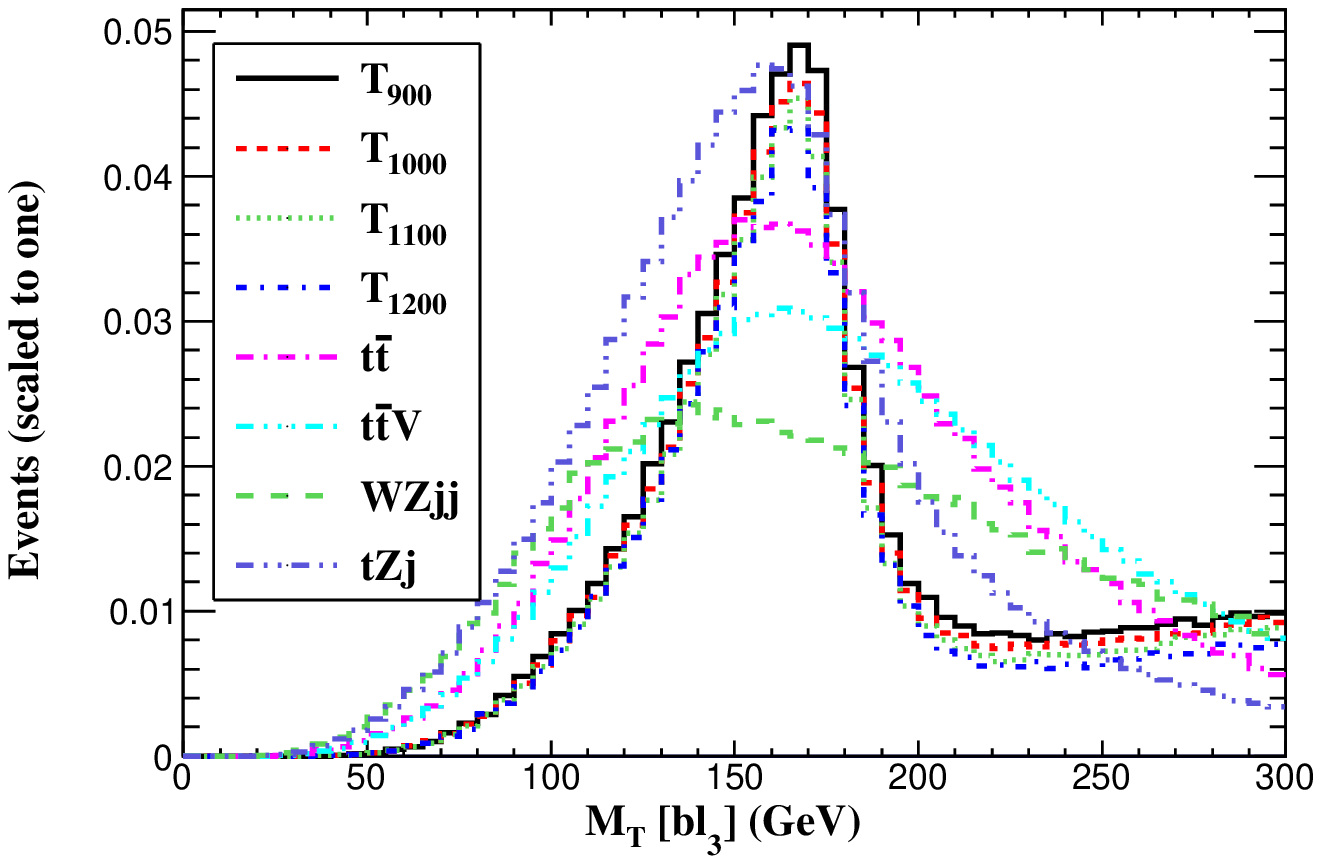}}
\centerline{\epsfxsize=9cm\epsffile{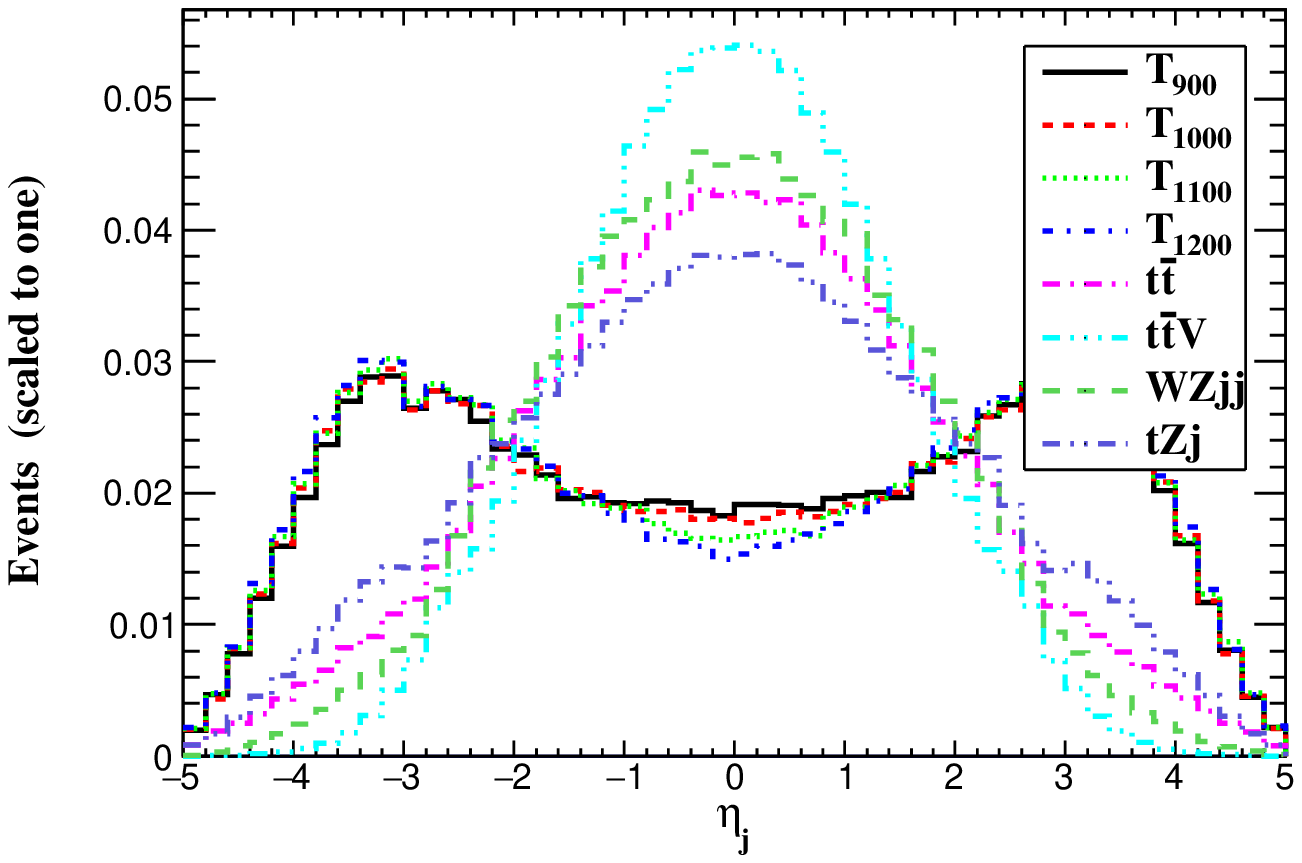}
\hspace{-1.0cm}\epsfxsize=9cm\epsffile{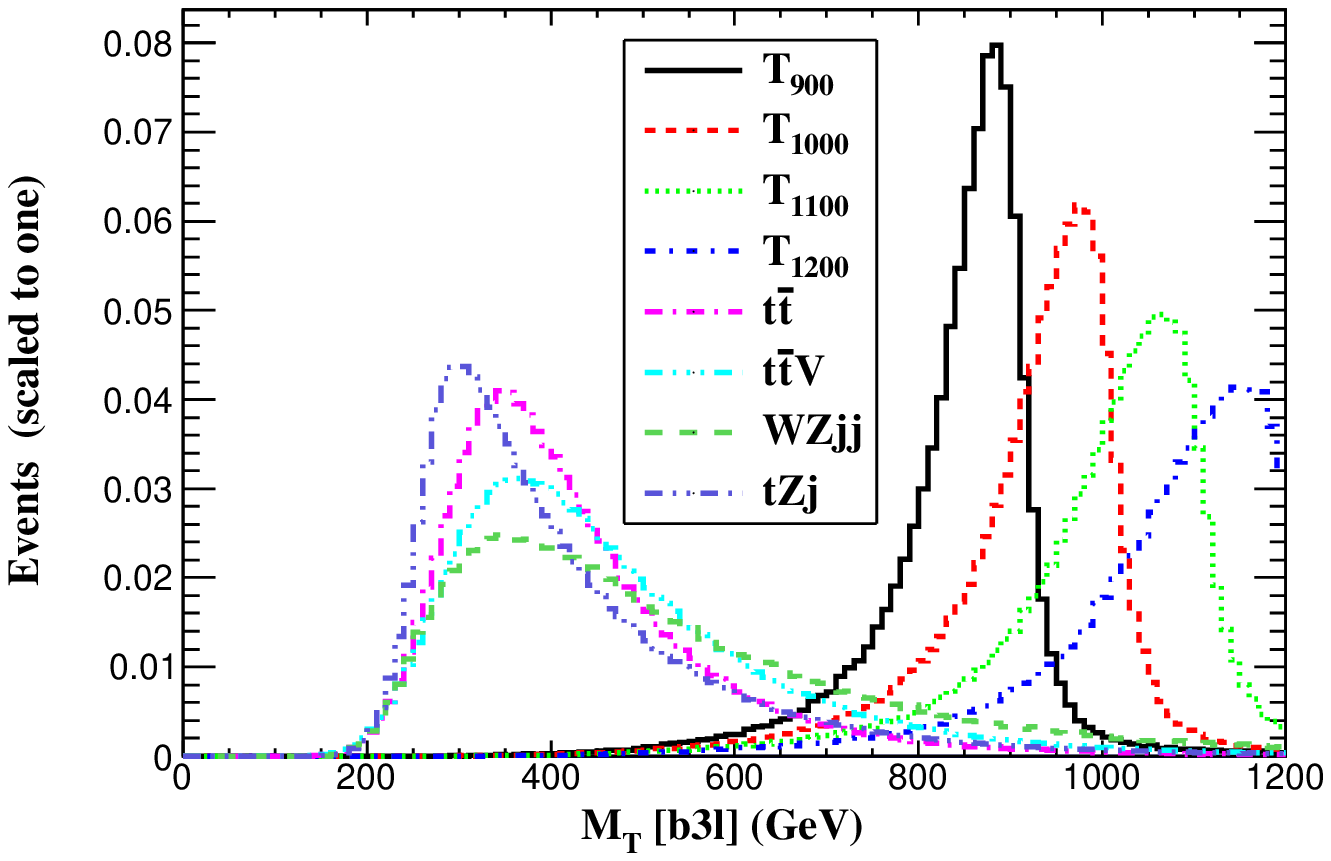}}
\caption{Normalized distributions for the signals and backgrounds}
\label{tz}
\end{center}
\end{figure}

Based on the kinematical distributions of the signal and backgrounds in Fig.~\ref{tz}, we furthermore impose the following cuts to get a high significance.
\begin{itemize}
\item
Cut-2: The transverse momenta of the leading and subleading leptons from the $Z$ boson are required to have $p_{T}^{\ell_{1}}> 150 \rm ~GeV$ and $p_{T}^{\ell_{2}}> 80 \rm ~GeV$, and the invariant mass of the $Z$ boson is required to have $|M(\ell_{1}\ell_{2})-m_{Z}|< 15 \rm ~GeV$.
\item
Cut-3: The transverse mass of the top quark is reconstructed as $140 \rm ~GeV <M_{T}(b\ell_{3})< 190 \rm ~GeV$.
\item
Cut-4: We require the light untagged jet to have $\mid\eta_{j}\mid > 2.4$.
\item
Cut-5: The transverse mass of the top partner decay products $3\ell b$ is required to have $M_{T}(b3\ell)> 800 \rm ~GeV$.
\end{itemize}

\begin{table}[htb]
\begin{center}
\caption{The cut flow of the cross sections (in fb) for four typical signals and the relevant
backgrounds at the 14 TeV LHC. Here we take the parameter as $g^{\ast}=0.2$\label{cutflow2}}
\vspace{0.2cm}
\begin{tabular}{|c|c|c|c|c|c|c|c|c|}
\hline
\multirow{2}{*}{Cuts}& \multicolumn{4}{c|}{Signal}&\multirow{2}{*}{$t\bar{t}$} &\multirow{2}{*}{$t\bar{t}V$} &\multirow{2}{*}{$WZjj$}
  &\multirow{2}{*}{$tZj$}\\ \cline{2-4}
 &{900~GeV} &{1000~GeV}&{1100~GeV}&{1200~GeV}&&&& \\  \cline{1-8}
\hline
Basic cuts& 0.49&0.33&0.21&0.087&14814&1.52&36.44&2.66 \\ \hline
Cut-1&0.068&0.036&0.02&0.008&0.58&0.16&0.84&0.53\\ \hline
Cut-2&0.046&0.024&0.013&0.0052&0.0096&0.011&0.12&0.029\\ \hline
Cut-3&0.041&0.021&0.012&0.0047&0.0048&0.005&0.039&0.023\\ \hline
Cut-4&0.023&0.012&0.0065&0.0027&0.001&$6.5\times 10^{-4}$&0.0072&0.0052\\ \hline
Cut-5&0.02&0.011&0.006&0.0026&$2.1\times 10^{-4}$&$9.2\times 10^{-5}$&$7.2\times 10^{-4}$&$9.3\times 10^{-4}$\\ \hline
\end{tabular} \end{center}\end{table}

We present the cross sections of four typical signal ($m_T=900, 1000, 1100, 1200$ GeV) and the relevant
backgrounds after imposing
the cuts in Table~\ref{cutflow2}.
From Table~\ref{cutflow2},
one can see that all the backgrounds are suppressed very efficiently after imposing the selections.
For the integrated luminosity $\pounds_{int}=3$ ab$^{-1}$, the number of events for total backgrounds after Cut-5 is found to be about six, while for the signal we obtain about 33 events for $m_T=1$ TeV and $g^{\ast}=0.2$.

\begin{figure}[htb]
\begin{center}
\centerline{\epsfxsize=9cm \epsffile{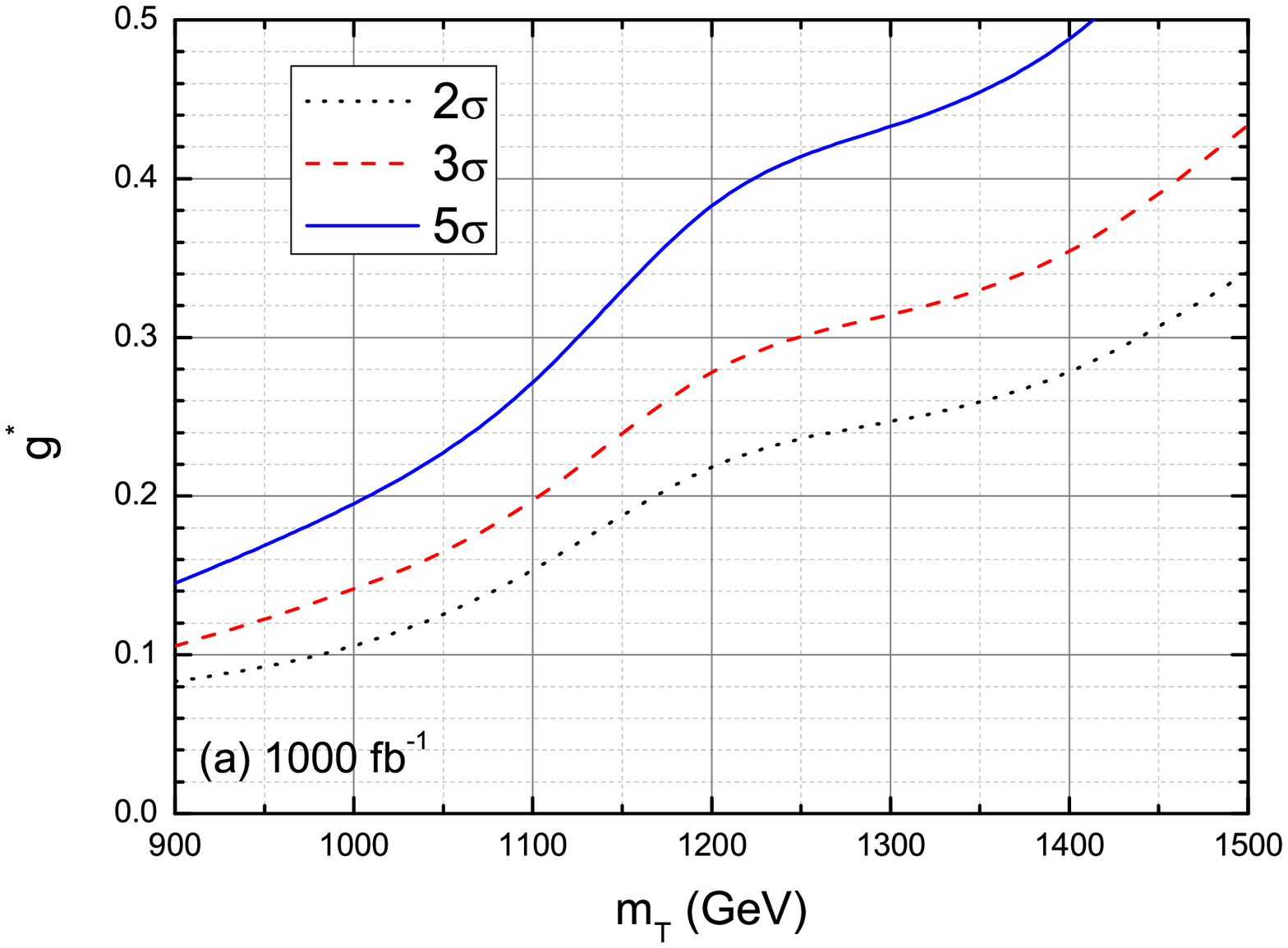}\epsfxsize=9cm \epsffile{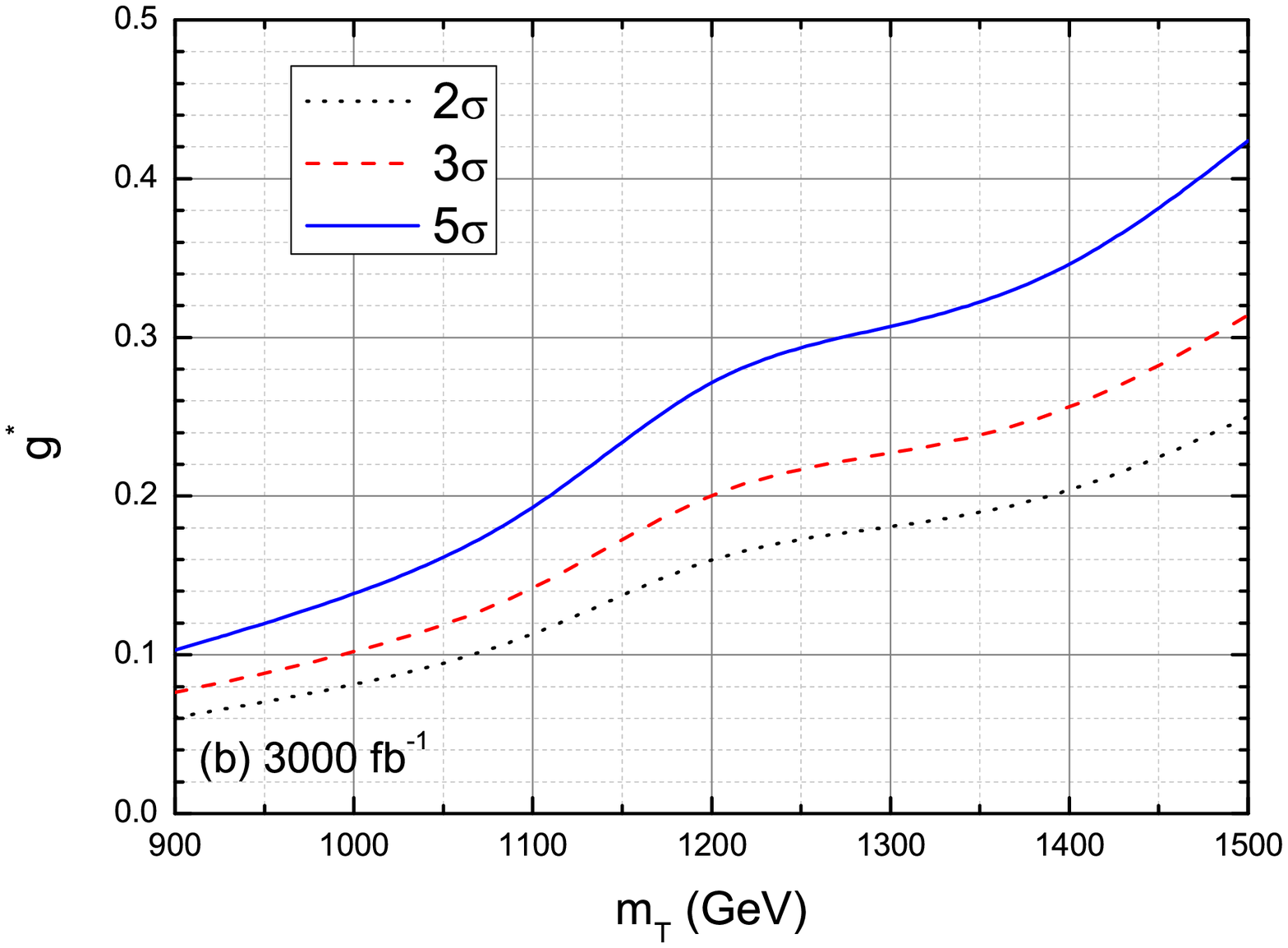}}
\caption{The $2\sigma$, $3\sigma$ and $5\sigma$ contour plots for the signal in $g^{\ast}-m_T$ at 14 TeV LHC with two typical values of integrated luminosity: (a) 1000 fb$^{-1}$, (b) 3000 fb$^{-1}$}
\label{ss-tz}
\end{center}
\end{figure}

In Fig.~\ref{ss-tz}, we plot the excluded $3\sigma$ and $5\sigma$ discovery reaches as a function of
$g^{\ast}$ and the top partner mass $m_T$ for two typical values of integrated
luminosity: 1000 and 3000 fb$^{-1}$. We can see that, for $m_T=1.0$ TeV,
the $5\sigma$ level discovery sensitivities of $g^{\ast}$ are respectively about
 0.14 with $\pounds_{int}=1000$ fb$^{-1}$ and 0.2 with $\pounds_{int}=3000$ fb$^{-1}$. On the other hand, the upper limits on the size
    of $g^{\ast}$ for $m_T=1.0~(1.4)$ TeV are, respectively, given as about $g^{\ast}\leq 0.11~(0.28)$  with
    $\pounds_{int}=1000$ fb$^{-1}$
     and $g^{\ast}\leq 0.14~(0.38)$ with $\pounds_{int}=3000$ fb$^{-1}$.

\begin{figure}[htb]
\begin{center}
\centerline{\epsfxsize=9cm \epsffile{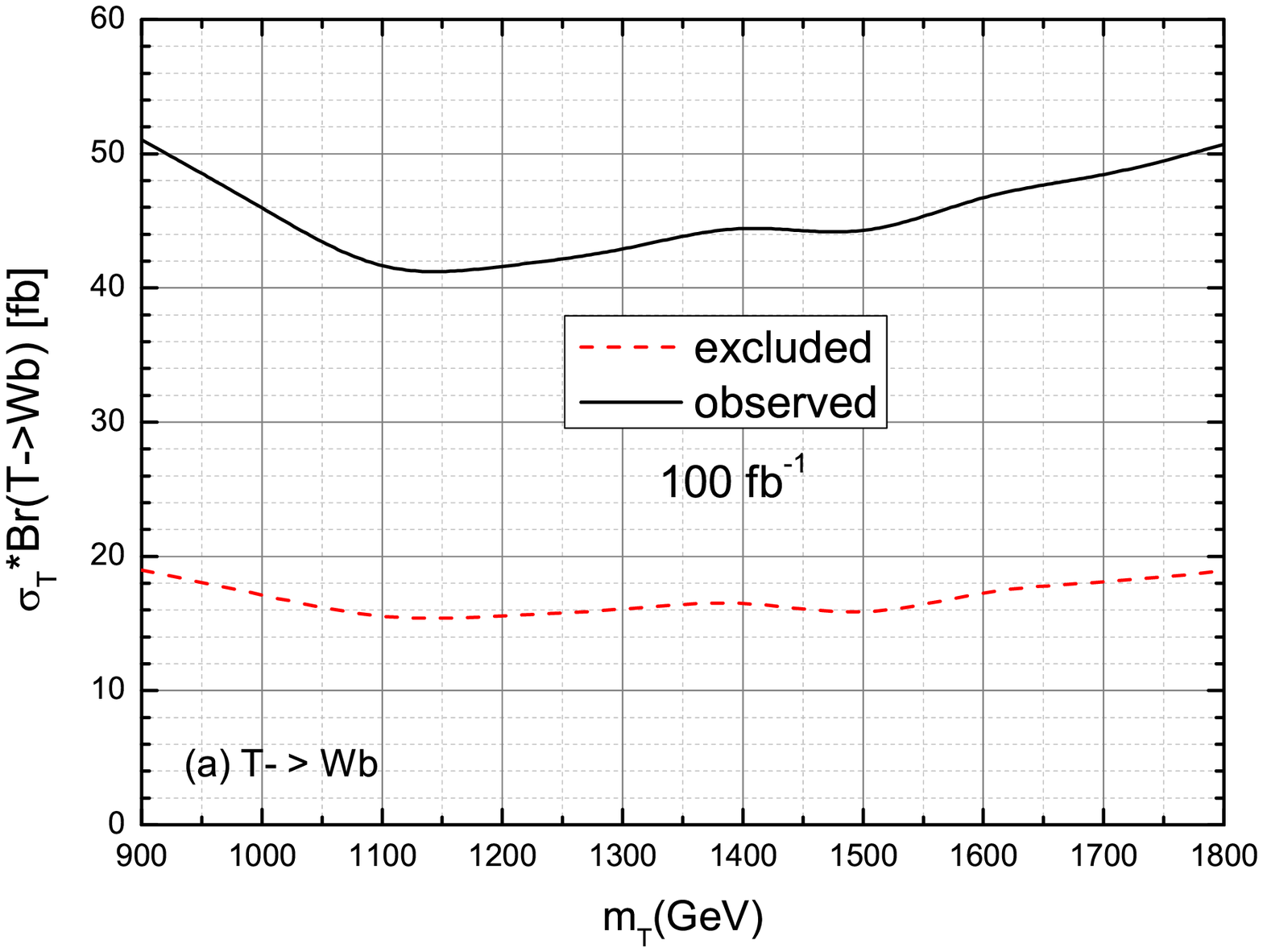}\epsfxsize=9cm \epsffile{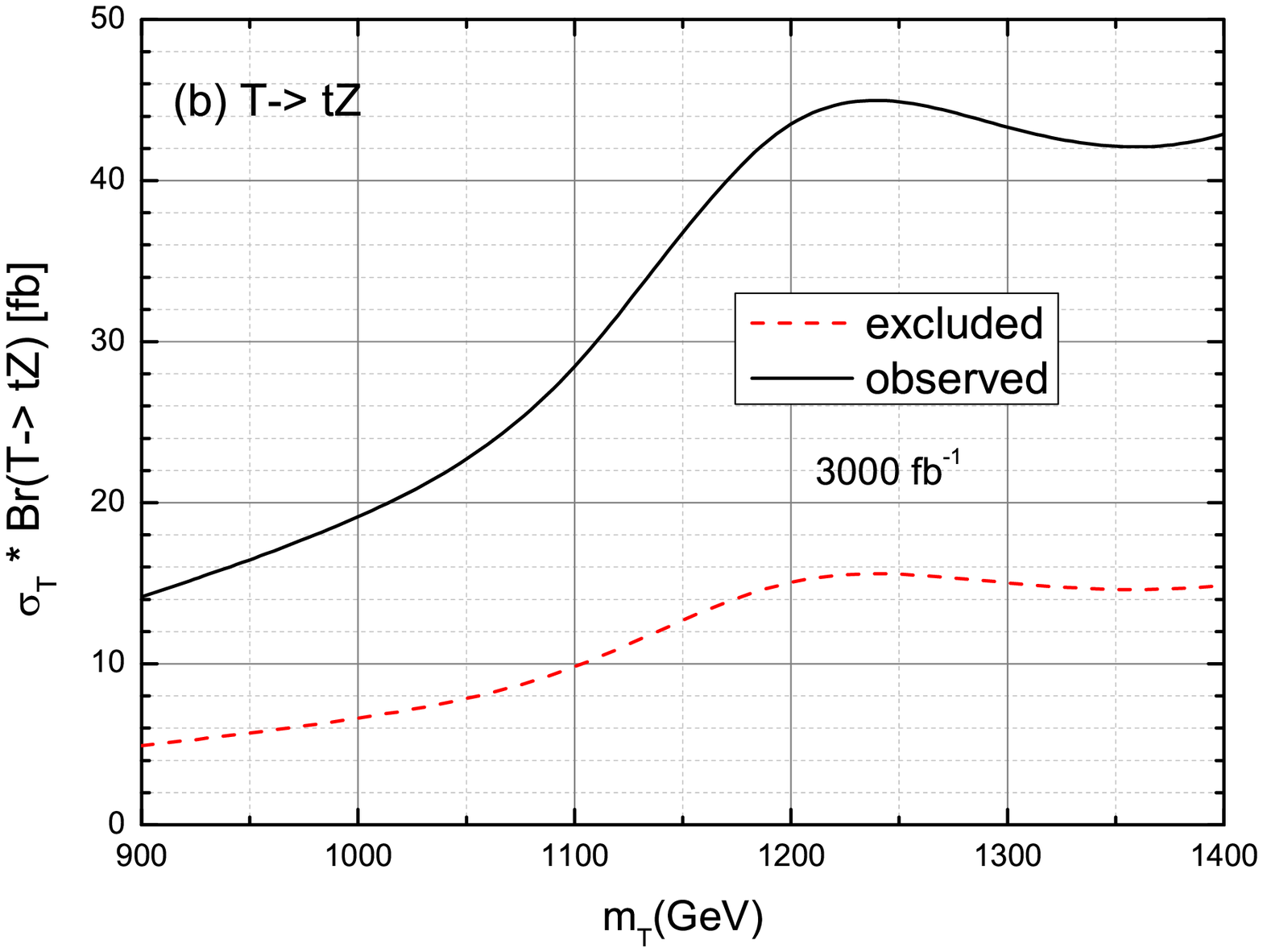}}
\caption{The excluded and observed cross section for the signal as a function of the vector-like top partner $m_T$ at 14 TeV LHC for (a) $T\to Wb$ and (b) $T\to tZ$ channels }
\label{CR}
\end{center}
\end{figure}

Note that our results are obtained from simulations of a simplified model implementation which in particular fixes the branching fractions between the different top partner decay channels, while the branching ratios can be altered in other general NP models including the $T$-quark. In Fig.~\ref{CR}, we plot their projected sensitivity in terms of the production cross section times branching ratio~($\sigma_T\ast Br$) for two decay channels. For the $T\to Wb$ channel, we find that the the single production cross sections of $\sigma_T\ast Br(T\to Wb)\sim40$-50 fb could be discovered at the 14 TeV LHC with 100 fb$^{-1}$ for $m_T\in [900, 1800]$ GeV, while the cross sections $\sim15$-20 fb will be excluded at the 14 TeV LHC with 100 fb$^{-1}$. For the $T\to tZ$ channel, we find that the the single production cross sections of $\sigma_T\ast Br(T\to tZ)\sim15$-45 fb could be discovered at the 14 TeV LHC with 3000 fb$^{-1}$ for $m_T\in [900, 1400]$ GeV, while the cross sections $\sim5$-15 fb will be excluded.

We can now
draw a comparison with other complementary studies for searches at the LHC run II
involving a singlet top partner. In Ref~\cite{NP1}, the authors show that a mass reconstruction
is possible within the $T\to th$ decay channel at $\sqrt{s }= 14$ TeV with
100 fb$^{-1}$ of integrated luminosity, proposing a search strategy optimised for two typical
top partner
mass points, namely $m_T= 800, 900$ GeV, and assuming $Br(T\to th)=100\%$. Furthermore, the authors
 in Ref.~\cite{NP2} designed a dedicated search
strategy for the leptonic $T\to bW$ decay channel at $\sqrt{s }= 14$ TeV with 30 fb$^{-1}$
of integrated luminosity, obtaining an expected exclusion reach
for masses up to 1.0 TeV, including both pair and single production.
For two typical top-partners masses $m_T=1.0~(1.5)$ TeV, the authors in Ref.~\cite{jhep-1604-014} studied
the search strategies of the single top-partner production with all the possible decay
modes (i.e., $tZ$, $th$ and $Wb$) at the LHC for $\sqrt{s }= 14$ TeV. The results show
that, for the specific model implementation discussed in Ref.~\cite{jhep-1604-014},
the production cross sections of $\sigma_T\sim$70-140~(30-65) fb for $m_T=1~(1.5)$ TeV, respectively, could be discovered at the LHC with 100 fb$^{-1}$. Similarly, the cross sections of $\sigma_T\sim$27-60~(13-24) fb for $m_T=1~(1.5)$ TeV, respectively, can be excluded.
Therefore, our analysis is competitive with the results of the
existing literature and represents a complementary candidate to search
for a possible singlet top partner.

\section{CONCLUSION}
In this paper, we have studied the prospects of observing the single $T$ production at the 14 TeV LHC in the $bW$ and $tZ$ decay channels. To illustrate our results, we adopt a simplified model including a $SU(2)_L$ singlet with charge $2/3$ with
only two free parameters, namely the $TWb$ coupling parameter $g^{\ast}$ and the top partner mass $m_T$. Since the single top partner production depends on the $TWb$
coupling parameter $g^{\ast}$ and the top partner mass $m_T$, the $2\sigma$ exclusion limits, $3\sigma$ evidence and the $5\sigma$ discovery reach in the parameter plane of $g^{\ast}-m_T$,
are obtained for various of integrated luminosity at the LHC Run II.
In the $T\to bW\to b\ell \nu$ decay channel, we rely on the large transverse momentum
of the $b$-jet, the lepton, and the forward nature of the light jet to suppress the backgrounds.
In the $T\to tZ$ decay channel, although the leptonic decay of the $Z$ entails a large suppression from the $Z$ leptonic branching ratio, the clean multilepton final state allows to strongly reduce the backgrounds and to reconstructed the top partner mass with high luminosity.

Even though we work in a simplified model including the singlet vector-like top partner, our results can also be mapped within the context of the specific models where the heavy $T$-quark only couplings to the third generation
of SM quarks, such as the minimal composite Higgs model~\cite{jhep-1304-004} and the littlest Higgs model with T-parity~\cite{lht}. We present a detailed
analysis of their projected sensitivity in terms of the production cross section times branching fraction for the relevant decay. At the 14 TeV LHC with 100 fb$^{-1}$, we find that the the single production cross sections of $\sigma_T*Br(T\to Wb)\sim 40$-50 fb could be discovered for $m_T\in [900, 1800]$ GeV, while the cross sections $\sim15$-20 fb will be excluded. For the $T\to tZ$ channel, we find that the the single production cross sections of $\sigma_T*Br(T\to tZ)\sim15$-45 fb could be discovered at the 14 TeV LHC with 3000 fb$^{-1}$ for $m_T\in [900, 1400]$ GeV, while the cross sections $\sim5$-15 fb will be excluded.
 We expect our analysis can represent a complementary candidate to pursue the search and mass measurement of a possible singlet top partner at the 14 TeV LHC.

\begin{acknowledgments}
This work is supported by the Joint Funds
of the National Natural Science Foundation of China (Grant No. U1304112),
the Foundation of He¡¯nan Educational Committee (Grant No. 2015GGJS-059)
 and the Foundation of Henan Institute of Science and Technology (Grant No. 2016ZD01).
\end{acknowledgments}


\end{document}